\documentclass[aps,prl,showpacs,preprint]{revtex4}
\usepackage{amssymb}
\usepackage{graphicx}
\usepackage{amsmath}
\usepackage{graphicx}

\input{tcilatex}

\begin{document}

\title{The spin-torque transistor}
\author{Gerrit E. W. Bauer}
\affiliation{Department of NanoScience, Delft University of Technology, 2628 CJ Delft,
The Netherlands}
\author{Arne Brataas}
\affiliation{Department of Physics, Norwegian University of Science and Technology,
N-7491 Trondheim, Norway}
\author{Yaroslav Tserkovnyak}
\affiliation{Harvard University, Lyman Laboratory of Physics, Cambridge, Massachusetts
02138 USA}
\author{Bart J. van Wees}
\affiliation{Department of Applied Physics and Materials Science Centre, University of
Groningen, Nijenborgh 4.13, 9747 AG Groningen, The Netherlands}

\begin{abstract}
A magnetoelectronic thin-film transistor is proposed that can display
negative differential resistance and gain. The working principle is the
modulation of the soure-drain current in a spin valve by the magnetization
of a third electrode, which is rotated by the spin-torque created by a
control spin-valve. The device can operate at room temperature, but in order
to be useful, ferromagnetic materials with polarizations close to unity are
required.
\end{abstract}

\date{\today }
\maketitle

Magnetoelectronic circuits differ from conventional ones by the use of
ferromagnetic metals. Electric currents depend on the relative orientation
of the magnetization vector of different magnetic elements, giving rise to
the giant magnetoresistance (GMR). The additional functionalities are useful
for sensing and data storage applications, like magnetic random access
memories (MRAMs) \cite{Parkin}. Several ideas on how to employ the spin
degree of freedom for other applications exist \cite{Devices,Wees}.

Here we pursue the \textquotedblleft spin-flip transistor\textquotedblright
, a three-terminal device consisting of an antiparallel spin-valve in which
the conducting channel is in contact with a ferromagnetic base \cite%
{Brataas00}. The source-drain current is modulated by the base magnetization
direction, since the latter affects the spin accumulation in the conducting
channel. It has been predicted \cite{Slon96,Berger96} and measured \cite{Rev}%
\ that the magnetization in spin valves can be switched by an electric
current. In Ref. \cite{Xia02} it was suggested to use the spin-flip
transistor as an MRAM element, in which the base magnetization is switched
by the spin-torque due to the induced spin accumulation. In the following,
we investigate the device parameters of the spin-flip transistor operated as
an amplifier by controlling the base magnetization by a second spin valve in
an integrated device that we call \textquotedblleft spin-torque
transistor\textquotedblright\ (Fig. 1). The lower part of this device
consists of$\ $source and drain contacts made from high-coercivity metallic
magnets with antiparallel magnetizations that are biased by an
electrochemical potential $\mu _{S}$. The source-drain electric current $%
I_{SD}$ induces a spin accumulation in the normal metal node \textit{N1}. We
attach an electrically floating base (or gate) electrode \textit{B, }which%
\textit{\ }is magnetically very soft and has good electric contact to 
\textit{N1}. When the magnetization angle $\theta $ is not $0$ or $\pi $ a
spin current flows into the base\ that decreases the spin accumulation and
increases $I_{SD}$ with $\theta $ up to $\pi /2$. On the other hand, the
spin accumulation in \textit{N1} exerts a torque on \textit{B }which strives
to lower $\theta .$ $\theta ,$ and thus $I_{SD}$ could be modulated, \textit{%
e.g}., by the \O rsted magnetic field generated electrically by the\
\textquotedblleft write line\textquotedblright\ of an MRAM element, but this
does not appear viable. We therefore propose the transistor in Fig. 1, which
integrates a second spin valve with magnetizations rotated by $\pi /2$ from
the lower one. An applied bias $\mu _{B}$ creates a another torque which
pulls the magnetization into the direction collinear to the upper contacts.
The base electrode then settles into a configuration at which both torques
cancel each other. A variation in $\mu _{B}$ then modulates $\theta $ and
consequently $I_{SD}.$ In the following we discuss the figures of merit of
the transistor action, \textit{viz}. the transconductance and the current
gain of this device.

For most transition metal based structures exchange splittings are large,
Fermi wavelengths short, and interfaces disordered. Electron propagation is
therefore diffuse and ferromagnetic (transverse spin) coherence lengths are
smaller than the mean-free path \cite{Zangwill}. In these limits
magnetoelectronic circuit theory is a convenient formalism \cite%
{Brataas00,Bauer02}. Spin-flip relaxation can be disregarded in the normal
metal node of small enough structures, since Al and Cu have spin-flip
diffusion lengths of the order of a micron \cite{Wees}. Spin-flip in the
source and drain electrodes can simply be included by taking their
magnetically active thickness as the smaller of the spin-flip diffusion
length and physical thickness. The base electrode is assumed to be
magnetically soft and the thickness is taken to be smaller than the
spin-flip diffusion length. These assumption are not necessary, since
magnetic anisotropies and spin-flip in the base can readily be taken into
account, but these complications only reduce the device performance and will
be treated elsewhere. The source-drain current dependence on the base
magnetization angle $\theta \ $then reads \cite{Brataas00}:%
\begin{equation}
I_{SD}\left( \theta \right) =\frac{e}{h}\frac{g_{S}\mu _{S}}{2}\frac{2\left(
g_{B}^{\uparrow \downarrow }+g_{S}\left( 1-p_{S}^{2}\right) \right)
g_{S}^{\uparrow \downarrow }+g_{B}^{\uparrow \downarrow }\left(
-2g_{S}^{\uparrow \downarrow }+g_{S}\left( 1-p_{S}^{2}\right) \right) \cos
^{2}\theta }{2\left( g_{S}+g_{B}^{\uparrow \downarrow }\right)
g_{S}^{\uparrow \downarrow }+g_{B}^{\uparrow \downarrow }\left(
g_{S}-2g_{S}^{\uparrow \downarrow }\right) \cos ^{2}\theta },  \label{sdi}
\end{equation}%
where $g_{S}=g_{S}^{\uparrow }+g_{S}^{\downarrow }$ and $p_{S}=\left(
g_{S}^{\uparrow }-g_{S}^{\downarrow }\right) /g_{S}$ are the normal
conductance and polarization of the source, and $g_{S}^{\uparrow \downarrow }
$ and $g_{B}^{\uparrow \downarrow }$ are the\ \textquotedblleft mixing
conductances\textquotedblright\ of the source and base contacts,
respectively. Drain and source contact conductances are taken to be
identical. All conductance parameters are in units of the conductance
quantum $e^{2}/h$, contain bulk and interface contributions \cite{Bauer02},
can be computed from first-principles and are taken to be real \cite{Xia02}.
The torque on the base magnetization created by the spin-accumulation is
proportional to the transverse spin-current \cite{Bauer02} into \textit{B}: 
\begin{equation}
L_{B}\left( \theta \right) =\frac{1}{2\pi }\frac{p_{S}g_{S}g_{S}^{\uparrow
\downarrow }g_{B}^{\uparrow \downarrow }\sin \theta \mu _{S}}{2\left(
g_{S}+g_{B}^{\uparrow \downarrow }\right) g_{S}^{\uparrow \downarrow
}+g_{B}^{\uparrow \downarrow }\left( g_{S}-2g_{S}^{\uparrow \downarrow
}\right) \cos ^{2}\theta },
\end{equation}%
A steady state with finite $\theta $ exists when $L_{B}\left( \theta \right) 
$ equals an external torque, either from an applied magnetic field, or a
spin accumulation from the upper side in Fig. 1. The differential
source-drain conductance $\tilde{G}_{SD}$ subject to the condition of a
constant external torque reads: 
\begin{eqnarray}
\tilde{G}_{SD} &\equiv &\left( \frac{\partial I_{SD}\left( \theta \right) }{%
\partial \mu _{S}}\right) _{L_{B}}=\frac{I_{SD}}{\mu _{S}}+\left( \frac{%
\partial I_{SD}}{\partial \theta }\right) _{\mu _{S}}\left( \frac{\partial
\theta }{\partial \mu _{S}}\right) _{L_{B}} \\
&=&\frac{I_{SD}}{\mu _{S}}-\left( \frac{\partial I_{SD}}{\partial \theta }%
\right) _{\mu _{S}}\frac{L_{B}\left( \theta \right) }{\mu _{S}\left( \frac{%
\partial L_{B}\left( \theta \right) }{\partial \theta }\right) _{\mu _{S}}},
\end{eqnarray}%
where the first term on the right hand sides is the derivative with respect
to $\mu _{S}$ for constant $\theta $ and the second term arises from the
source-drain bias dependence of $\theta .$ The general equations are
unwieldy and not transparent. The most important parameter turns out to be
the spin-polarization $p_{S}\ $of the source and drain contacts. We
therefore choose a model system with $p_{S}$ variable, but other parameters
fixed for convenience, \textit{viz}. the same parameters for both spin-flip
transistors and $g_{B}^{\uparrow \downarrow }=g_{S}^{\uparrow \downarrow
}=g_{S},$ which holds approximately for metallic interfaces with identical
cross sections \cite{Xia02}. We find that 
\begin{equation}
\tilde{G}_{SD}=\frac{e^{2}}{h}\frac{g_{S}}{2}\left( 1-p_{S}^{2}\frac{2+\cos
^{2}\theta +\frac{4\sin ^{2}\theta }{2-\cos ^{2}\theta }}{4-\cos ^{2}\theta }%
\right) 
\end{equation}%
may become negative, since an increased source-drain bias tends to rotate
the angle to smaller values, thus reducing the source-drain current. At the
sign change of $\tilde{G}_{SD},$ the output impedance of the spin valve
becomes infinite, which can be useful for device applications.

We now demonstrate that it is attractive to modulate $I_{SD}$ by the
spin-transfer effect \cite{Slon96,Berger96, Waintal00}. In contrast to work
in the literature that was focussed on magnetization reversal by large
currents \cite{Rev}, we envisage controlled rotations by small voltages. The
base is supposed to be highly resistive, consisting of a magnetic insulator,
or, alternatively, of two magnetically coupled ultrathin magnetic films
separated by a thin insulator. The device might be realized in the lateral
thin-film geometry by van Wees \textit{c.s}. \cite{Wees}, using a soft
magnet with a circular disk shape for the base, sandwiched in a cross
configuration of normal metal films with ferromagnetic contacts. The device
characteristics can be computed for the complete parameter space by circuit
theory, but the important features are retained by proceeding as above and
also assuming the same parameters for the upper and lower sections. The
stationary state of the biased spin-transfer transistor is described by the
angle $\theta _{0}$ at which the two torques on the base magnet cancel each
other. For the present model this is the solution of the transcendental
equation 
\begin{equation}
\frac{\mu _{B}}{\mu _{S}}=\frac{7+\cos 2\theta _{0}}{7-\cos 2\theta _{0}}%
\tan \theta _{0}.
\end{equation}%
The calculated source-drain differential conductance (now without tilde) has
to be computed now under condition of constant $\mu _{B}$ rather then a
constant torque 
\begin{equation}
G_{SD}\equiv \left( \frac{\partial I_{SD}\left( \theta \right) }{\partial
\mu _{S}}\right) _{\mu _{B}}=\frac{I_{SD}}{\mu _{S}}+\left( \frac{\partial
I_{SD}}{\partial \theta }\right) _{\mu _{S}}\left( \frac{\partial \theta }{%
\partial \mu _{S}}\right) _{\mu _{B}}
\end{equation}%
which is plotted as a function of $\mu _{S}$ and polarization $p_{S}$ in
Fig. 2. Note that with increasing $p_{S}$ strong non-linearities develop,
which for large polarizations lead to a zero and negative differential
resistance at $\mu _{B}\approx \mu _{S}.$ The physical reason is the
competition between the Ohmic current, which for constant resistance
increases with the bias, and the increasing torque, which at constant $\mu
_{B}$ decreases the current, as noted above.

The differential transconductance measures the increase of the source-drain
current (at constant $\mu _{S}$) induced by an increased chemical potential
of the base electrode $T\left( \theta \right) \equiv \left( \partial
I_{SD}\left( \theta \right) /\partial \mu _{B}\right) _{\mu _{S}}.$ We focus
discussion here on the differential current gain, \textit{i.e}. the ratio
between differential transconductance and channel conductance $\Gamma
=T/G_{SD},$ as a representative figure of merit. In the regime $\mu _{B}\ll
\mu _{S}$ and thus small $\theta _{0}\rightarrow 3\mu _{B}/\left( 4\mu
_{S}\right) ,$ the current gain$\ $becomes 
\begin{equation}
\lim_{\mu _{B}\rightarrow 0}\Gamma =\frac{\frac{1}{2}\theta _{0}}{\frac{%
1-p_{S}^{2}}{1+p_{S}^{2}}-\frac{1}{3}\theta _{0}^{2}}.
\end{equation}%
For small polarizations the $\theta _{0}^{2}$ in the denominator may be
disregarded and $\Gamma \sim \theta _{0},$ thus is proportional to the
control potential $\mu _{B}.$ When the polarization is close to unity,
however, we see that $\Gamma $ becomes singular at small angles and changes
sign. This behavior reflects the negative differential resistance found
above for $\mu _{B}$ and $\theta $. For complete polarization $\left(
p_{S}=1\right) $ $\Gamma =-3/\left( 2\theta _{0}\right) .$ For polarizations
(slightly) smaller than unity we may tune the transistor close to the
optimal operation point of infinite output impedance 
\begin{equation}
\theta _{0,c}=\sqrt{3\frac{1-p_{S}^{2}}{1+p_{S}^{2}}}
\end{equation}%
at which $\Gamma \sim \left( \theta _{0}-\theta _{0,c}\right) ^{-1}.$

The working principle of this spin-transfer transistor is entirely
semiclassical, thus robust against, for example, elevated temperatures. The
derivations assumed absence of phase coherence and electron correlation, but
the physics most likely survives their presence. The base contact is
preferably a magnetic insulator or contains a thin insulating barrier (F%
\TEXTsymbol{\vert}I\TEXTsymbol{\vert}F), but the contact to the normal metal
should be good (for a large mixing conductance). Tunnel junctions may be
used for the source-drain contacts, but this will slow down the response
time. It should be kept in mind as well that the dwell time of electrons in
the device must be larger than the spin-flip relaxation time. The basic
physics, such as the non-linearity of the source-drain conductance in Fig.
2, should be observable for conventional ferromagnetic materials. Large
current gains exist for incomplete polarization close to unity of the source
and drain ferromagnets, but at the cost of non-zero \textquotedblleft
off\textquotedblright\ currents. A useful device should therefore be
fabricated with (nearly) half-metallic ferromagnets for sources and drains.
As base magnet, a thin film of any soft ferromagnet is appropriate as long
as it is thicker than the ferromagnetic (transverse spin) coherence length,
but not too thick in order to keep the response to torques fast. We
recommend a couple of monolayers of permalloy on both sides of a very thin
alumina barrier.

In conclusion, we propose a robust magnetoelectronic three terminal device
which controls charge currents via the spin-transfer effect. It can be
fabricated from metallic thin films in a lateral geometry, but its
usefulness will derive from the availability of highly polarized
(half-metallic) ferromagnets.

We would like to thank Prof. G. G\"{u}ntherodt for asking the question about
the gain of the spin-flip transistor. We acknowledge discussions with Paul
Kelly, Alex Kovalev, and Yuli Nazarov, as well as support by FOM, NSF Grant
DMR 02-33773 and the NEDO joint research program \textquotedblleft
Nano-Scale Magnetoelectronics\textquotedblright .

\begin{figure}
[ptb]
\begin{center}
\includegraphics[
]%
{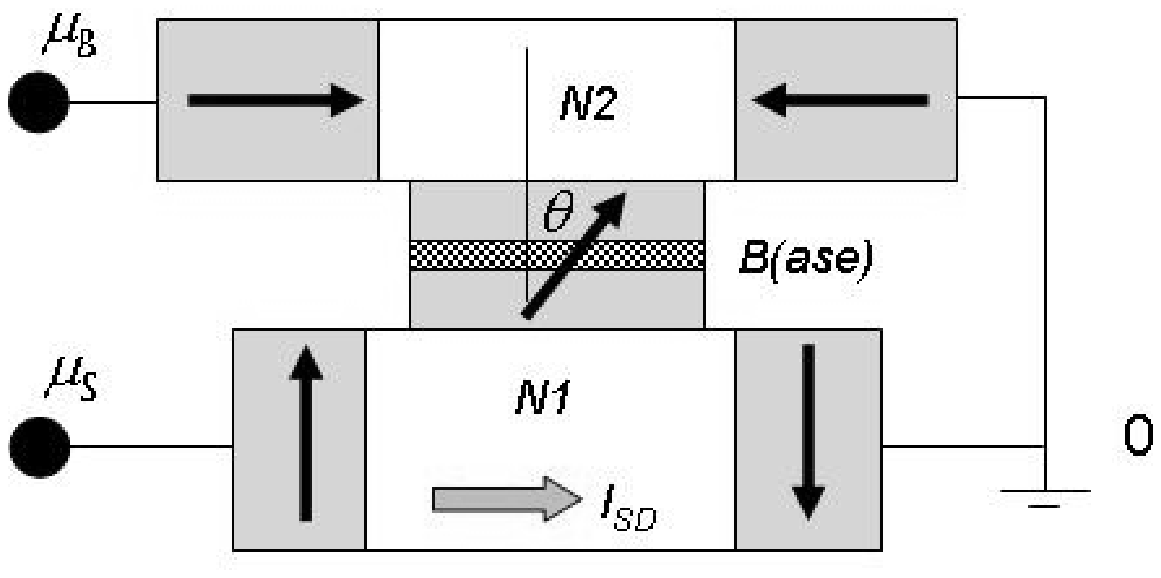}%
\caption{ Schematic sketch of the
spin-transfer transistor consisting of two spin-flip transistors with a
common base contact \textit{B} and source drain contact magnetizations which
are rotated by $90%
{{}^\circ}%
$. The magnetization direction of the base \textit{B} is controlled by the
chemical potentials $\protect\mu _{B}$ and $\protect\mu _{S}.$}%
\end{center}
\end{figure}
\begin{figure}
[ptb]
\begin{center}
\includegraphics[
height=7.7326cm,
width=10.4cm
]%
{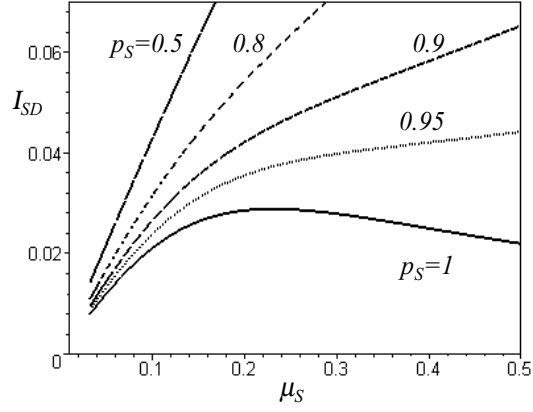}%
\caption{ Source-drain current Eq. (1) of
the spin-transfer transistor, divided by the contact conductance $%
e^{2}g_{S}/h$ , \textit{i.e}. in (voltage) units of $\protect\mu _{S}/e,$ as
a function of $\protect\mu _{S}$ and the polarization $p_{S}$ of the source
and drain contacts. A constant $\protect\mu _{B}=0.2$ (in the same units as $%
\protect\mu _{S}$) is applied.}%
\end{center}
\end{figure}

\end{document}